\documentstyle[prl,aps,multicol,epsfig]{revtex}
\begin{document}
\newcommand{\fig}[2]{\epsfxsize=#1\epsfbox{#2}}
\author{T. Giamarchi}
\address{Laboratoire de Physique des Solides, Universit{\'e} Paris-Sud,\\
B{\^a}t. 510, 91405 Orsay, France\cite{junk}}
\author{P. Le Doussal}
\address{CNRS-Laboratoire de Physique Th\'eorique de l'Ecole\\
Normale Sup\'erieure, 24 rue Lhomond, F-75231 Paris\cite{frad}}
\title{Phase diagrams of flux lattices with disorder}
\date{\today}
\maketitle

\begin{abstract}
We review the prediction, made in a previous work [Phys. Rev. B 
{\bf 52} (1995)],
that the phase diagram of type II superconductors
consists of a topologically ordered Bragg glass phase at low fields
undergoing a transition at higher fields into a
vortex glass or a liquid. We estimate the position of the
phase boundary using a Lindemann criterion.
We find that the proposed phenomenology 
is compatible with recent experiments
on superconductors.
\end{abstract}

\pacs{to be added}
\begin{multicols}{2}

\narrowtext

It is remarkable that after a decade of experimental and theoretical
efforts, the phase diagram of type II superconductors
in a field is far from being completely elucidated
\cite{blatter_vortex_review}. Stimulated by the
discovery of the high Tc materials, a reexamination of the mean-field phase
diagram unravelled two main new phenomena.
First, it was realized \cite{nelson_melting,subdo_melting},
and observed \cite{melting_experiments} that due to enhanced thermal
fluctuations the Abrikosov lattice melts
well below $H_{c2}$ into a to a flux liquid.
Secondly, it was argued that in the solid phase,
pointlike disorder could produce a glassy state with barriers $U(j)$ diverging
at small $j$, and thus
characterized by the true vanishing of the linear resistivity
even at finite temperature
\cite{fisher_vortexglass_short,feigelman_collective}.
This was a significant departure from traditional models of thermally assisted
flux flow, which assumed {\it finite} barriers between pinned states.
A precursor sign of an instability towards a glass was also found
in the flux liquid \cite{nelson_ledoussal_liquid}. Both for the
technological applications of high-$T_c$ materials and from a purely theoretical
point of vue, the understanding of the detailed properties of such a glassy
phase is of paramount importance.

Two main phenomenological theories have been put forward to describe
this glassy phase and to account for some of its
properties observed in early experiments, mainly the observed continuous transition
from the glass to the liquid and the giant thermal creep. The first approach
is based on the gauge glass model
\cite{fisher_vortexglass_short,fisher_vortexglass_long},
and assumes a complete destruction
of the Abrikosov lattice. The second approach retains the
the elastic lattice structure at small scale \cite{feigelman_collective}.
Although different in nature, both theories agreed that the disorder
essential to produce the glassy low temperature phase and the vanishing of
the linear resistivity, was also destroying at large scales the perfect flux lattice
existing in mean field theory.
The low temperature phase was therefore generally expected
to be a topologically disordered phase, lacking translational order.
Several calculations supported this point of view. Elastic
theory predicted at best a stretched exponential decay
of translational order
\cite{feigelman_collective,chudnovsky,bouchaud_variational_global}
(i.e. a power law grow of displacements) and general 
arguments tended to prove that
disorder would always favor the presence of 
dislocations \cite{fisher_vortexglass_long}.
The vortex lattice seemed to be buried for good.

\begin{figure}
\label{figure1}
\centerline{\epsfig{file=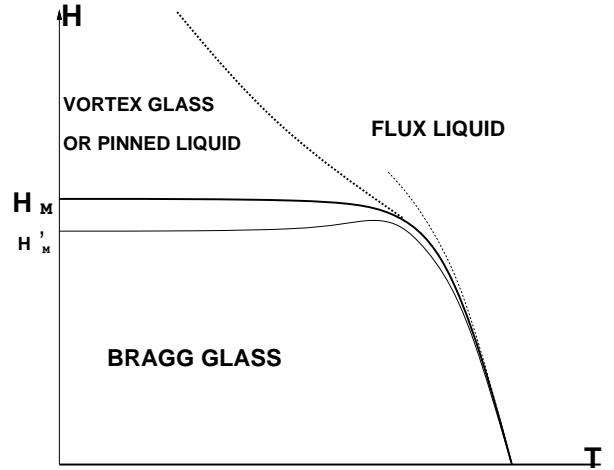,width=8cm}}
\vspace{0.5cm}
\caption{The stability region of the Bragg glass phase in the magnetic
field
$H$, temperature $T$ plane is shown schematically (thick solid line).
Upon increasing disorder the
region shrinks as indicated by the thin solid line (see text).
The melting line of the pure system is shown as a dotted line, and
the vortex glass transition line (or crossover to the pinned liquid)
is shown thick dotted.}
\end{figure}

A few points were not naturally fitting in the framework
of these theories. Experimentally,
a first order transition between the glass phase and the liquid
was observed at low fields \cite{charalambous_melting_rc,safar_tricritical_prl}
rather than the predicted continuous transition,
observed at high fields. Also, decoration experiments of
the flux lattice at very low fields (60 G) in several materials
were showing remarkably large regions
free of dislocations \cite{grier_decoration}.
On the side of theory, old calculations on the
related disordered elastic random field  XY model \cite{villain_cosine_realrg}
as well as more recent scaling arguments for the vortex lattice
\cite{nattermann_pinning} suggested, within a purely
elastic decription, a slower, logarithmic, growth of deformations.

In a recent work we have obtained the first quantitative theory of the
elastic vortex lattice
\cite{giamarchi_vortex_short,giamarchi_vortex_long} in presence 
of point disorder.
Contrarily to
previous approaches, it provides a description valid at all scales and 
demonstrates
that while disorder produces algebraic growth of displacements at short 
length scales,
periodicity takes over at large scales and results in a
decay of translational order {\it at most algebraic} \cite{footnote1}.
One striking prediction is thus the existence of a glass phase with
Bragg diffraction peaks !
This result was derived within an elastic theory, assuming the absence of
dislocations. However, the very result of our calculation, i.e
that quasi long range order survives, led us to advocate that dislocations
would be much less relevant than commonly assumed \cite{giamarchi_vortex_short}.
The alleged importance of dislocations in a disordered system
\cite{fisher_vortexglass_long} made it mandatory to further investigate
carefully whether dislocations would indeed modify the above
result. The striking result \cite{giamarchi_vortex_long} that we found
based on energy arguments is that dislocations are {\it not favorable}
for weak disorder in $d=3$. This implies self-consistently the existence
of a thermodynamic glass phase, as far as energy and very low
current transport properties are
concerned, retaining a nearly perfect (i.e. algebraic) translational order.
Since this phase exhibits Bragg
peaks very much like a perfect lattice it was christened the ``Bragg
glass''. Because it retains a ``lattice'' structure
and Bragg peaks, this glass phase is radically different from the vortex
glass picture based on a random gauge model. In particular,
since such a phase is nearly as good as perfect lattice as far
as translational order is concerned, it is natural to expect it to melt
through a first order phase transition. We proposed
\cite{giamarchi_vortex_long} that the phase seen experimentally
at low fields was in fact the Bragg glass, solving the
apparent impossibility of a pinned solid.
This allowed to account naturally for the first-order transition
and the decoration experiments. Our prediction \cite{giamarchi_vortex_long}
that a new phase
without topological defects, should be stable
at weak disorder received subsequent further support both from numerical
simulations \cite{gingras_dislocations_numerics,ryu_diagphas_numerics}
and from analytical calculations in a layered geometry
\cite{kierfeld_bglass_layered,carpentier_bglass_layered}.

Once the existence of a weak disorder/low field Bragg glass
phase is established the question arises of 
determining its limits of stability and phase boundaries.
The Bragg glass phase should be stable as a self-consistent
solution in the elastic limit, i.e as long as 
$R_a \gtrsim a$ \cite{giamarchi_vortex_long}. 
This condition is violated when the
field is increased and we proposed in \cite{giamarchi_vortex_long}
that upon raising the field the Bragg glass should
undergo a transition into another phase, which could
be a pinned liquid or another glass (vortex glass).
A natural possibility then was that the
critical point occuring on the melting line \cite{safar_tricritical_prl}
was the end point of the transition line between the Bragg glass at low fields
and a topologically disordered glassy phase (or a strongly
pinned liquid) at higher field. 
We pointed out that the fact that this point can
be lowered by adding impurities was a hint that it was
related to this transition. Such a field-driven
transition corresponds to the destruction of the Bragg glass
by proliferation of topological defects
upon raising the field, which is equivalent to increasing the
effective disorder. The other transition from the thermal liquid
into the putative superconducting state at higher fields is
presumably continuous. The topology of the phase diagram
proposed in \cite{giamarchi_vortex_long} 
is as depicted in Fig. 1.

Several recent experiments can be interpreted to confirm the picture
proposed in \cite{giamarchi_vortex_long}.
Neutron experiments can be naturally interpreted in term of
the Bragg glass \cite{yaron_neutrons,giamarchi_comment_neutrons}.
In BSCCO neutron peaks are observed at low fields and disappear upon
raising the field \cite{forgan}. The phase diagram of BSCCO
has been recently explored in details by overcoming spurious effects
due to geometrical barriers \cite{melting_bscco}. It can increasingly 
be interpreted as a confirmation of our theory, the so-called second magnetization
peak line being the candidate for the predicted field driven transition.
Since our proposal this line has been investigated in more details in BSCCO
\cite{khaykovich_zeldov} and found to be relatively temperature independent
at lower temperatures and to be shifted downwards upon increase of point disorder
through electron irradiation \cite{khaykovich_diagphas_bisco}.
Also, similar types of phase diagrams are observed in a variety of materials,
including YBCO \cite{safar_tricritical_prl,safar_transport_tricritical}.
The fact that a controlled increase of point disorder through electron irradiation
shifts the transition line to lower fields \cite{kwok_irradiations}
is a strong indication that our picture is relevant in these
materials as well.

In the present paper we follow on the theory exposed in
\onlinecite{giamarchi_vortex_long}. We make more quantitative estimates
of the phase diagram depicted in Fig. 1 using a
generalized Lindemann criterion. We also explore in more details
some experimental consequences of our theory.

Let us consider a vortex lattice system in presence of disorder. We can
model the vortex lattice by stacks of coupled planes.
The system is therefore described by layers of two dimensional
triangular lattices of vortices.
We denote by
$R_i$ the equilibrium position of the vortices in the absence of
disorder, labeled by an integer
$i$, in the $xy$ plane, and by $u(R_i,z)$ their in-plane displacements
which are two dimensional
vectors, (the vortex can only move within the plane).
$z$ is the coordinate perpendicular to the planes and along the
magnetic field and $x=(r,z)$.
The total energy is:
\begin{eqnarray} \label{vordep}
&& H = \frac12 \int d^2r dz [(c_{11}-c_{66})(\partial_\alpha u_\alpha)^2
+c_{66} (\partial_\alpha u_\beta)^2 + \nonumber \\
&& c_{44}
(\partial_z u_\alpha)^2]  + \int d^2r dz V(r,z) \rho(r,z)
\end{eqnarray}
where the density of vortex lines is simply defined by
$\rho(x) = \sum_i \delta^{(2)}(r - R_i -u(R_i,z))$.
The last term in (\ref{vordep}) is the coupling to disorder.
In the limit where many weak impurities act collectively on a vortex,
point disorder can be modelled by
a gaussian random potential $V(x)$ with correlations:
$\overline{V(x)V(x')}=\Delta(r-r')\delta(z-z')$ where $\Delta(r)$
is a short range function \cite{footnote2}
of range $\xi$ (the superconducting
coherence length)
\begin{equation}
\Delta(r) = d ~ U_p^2 e^{-r^2/\xi^2}
\end{equation}
where $d$ is the distance between layers and $U_p$ a typical pinning
energy per unit length along $z$.

In the high-$T_c$ Abrikosov lattice, one has in principle to use
non local elasticity, and the calculation along the lines of
\onlinecite{giamarchi_vortex_long} can be done simply by
using the complete known wavevector dependent
expressions \cite{subdo_melting,glazman_koshelev_decoupling}
of the elastic constants in (\ref{vordep}). Since we
are only interested in nearest neighbor
correlations and want to obtain only an order of magnitude of the scales
involved we use simple constant elastic moduli.
The physical properties of (\ref{vordep}) were examined in details in
\onlinecite{giamarchi_vortex_short,giamarchi_vortex_long} and we just
recall here the results needed for the phase diagram. The mean
squared relative displacements of two vortices separated
by a distance $r$ is:
\begin{equation} \label{linmelt}
B(r) = \overline{\langle (u(0,0) - u(r,0))^2 \rangle}
\end{equation}
where $\langle\rangle$ denotes the thermal average, whereas
$\overline{...}$ is the disorder average. From
$B(r)$ one defines two length scales $R_c$ and $R_a$ in the
$xy$ plane (and similarly $L_c$ and $L_a$ along $z$) such that
$B_{dis}(R_c)\sim \max(\xi ^2, \langle u^2 \rangle_T)$
(see below) and $B(R_a)\sim a^2$ respectively. $R_c$ is the
Larkin-Ovchinikov pinning length \cite{larkin_ovchinnikov_pinning}
directly related to the critical current,
whereas $R_a$ is the scale at which one enters the asymptotic
regime with a logarithmic growth of the displacements. The model
(\ref{vordep}) leads to the Bragg glass phase with quasi long
range translational order.

To determine the region of stability of the Bragg glass phase, we follow
the arguments proposed in \onlinecite{giamarchi_vortex_long} that
the elastic structure will become unstable when the displacement between two
neighbors becomes of order of the lattice spacing $a$, i.e
\begin{equation}
B(r=a) \sim a^2
\end{equation}
To be more quantitative, one can introduce, as for the normal thermal
melting a Lindermann constant $c_L$ and take for the
criterion of stability of the Bragg glass phase
\begin{equation} \label{criterion}
B(r=a) = \overline{\langle (u(0,0) - u(a,0))^2 \rangle}= c_L^2 a^2
\end{equation}
$c_L$, the Lindemann constant is usually of the order of $c_L\sim 0.1-0.2$
in usual melting and we make here the assumption
that $c_L$ is indeed a constant independent of
the field.

 From (\ref{criterion}) one sees that both disorder and thermal
fluctuations act together to increase the displacements. In fact
the formula (3.18) of \onlinecite{giamarchi_vortex_short,giamarchi_vortex_long}
shows that $B(r=a)$ splits naturally in two parts $B(r=a) \approx
2 \langle u^2 \rangle_T + B_{dis}(r=a)$. One immediate
consequence of (\ref{criterion}) is therefore that the melting
line should be pushed {\it downwards} in the presence of point like
disorder.
In fact the Bragg glass can disappear in two ways:
(i) if the temperature is raised, it will melt to a liquid phase
(ii) if the field is raised,
which amounts to vary the effective disorder in the system, the system
can become so disordered even at short length scales that dislocations
will appear. (\ref{criterion}) gives thus the limit of stability of
the BG phase in the $H-T$ plane.
Although the complete ``melting'' curve can be computed using the
formulas for $B(r)$ obtained in
\onlinecite{giamarchi_vortex_short,giamarchi_vortex_long}, such a
calculation is tedious and offers little insight. We therefore study
mainly here the two limits of low temperature where the transition is
mainly field-driven, and for temperature close to the melting curve in
the absence of disorder.

If the temperature is close to the pure melting line, $B(r=a)$ is
dominated by the thermal fluctuations. Since for weak disorder $R_a \gg
a$ the disorder induced displacements are negligeable at the scale of
nearest neighbors and one can compute (\ref{criterion}) using thermal
fluctuations only. One then easily recovers the pure melting line
\begin{equation}
T_m \approx 4 a^3 \sqrt{c_{66} c_{44}} c_L^2
\end{equation}
Disorder effects will push the melting line slightly down, but effects
should be negligible at low field for which the effective disorder is
small enough.
Upon increasing the field disorder induced displacements will increase,
forcing the transition line defined by (\ref{criterion})
to go down to zero temperature at a finite field $H_M$. The scale at which
disorder dominates can easily be obtained by looking at zero
temperature. To obtain a reliable order of magnitude of the 
``disorder-induced melting''
field $H_M$,
it is necessary to know the precise $B(r)$ in the presence of
disorder and not only its asymptotic forms. Fortunately such a
calculation was performed in
\cite{giamarchi_vortex_short,giamarchi_vortex_long}.
Using formula (4.18) of Ref. \onlinecite{giamarchi_vortex_long} one gets
\begin{equation}
B(r) = \frac{a^2}{\pi^2} \tilde{b}(r/R_a)
\end{equation}
For $r=R_a$ one has from \onlinecite{giamarchi_vortex_long}
that $\tilde{b} \approx 1$ while
for $r< R_a$ one is in the random manifold regime
and one can approximate
$B(r)\simeq  \frac{a^2}{\pi^2} (r/R_a)^{1/3}$. From the solution of
\onlinecite{giamarchi_vortex_long} we know that the above formula is
{\it quantitatively} correct, and not only asymptotics. Using
(\ref{criterion}) one finds that
\begin{equation} \label{lera}
a/R_a = (\pi c_L)^6
\end{equation}
Using $c_L = 0.12$ gives $R_a \sim 350 a$. Thus the 
transition occurs well before the asymptotic regime.
We will find that it does occur (e.g in BSCCO)
indeed deep into the random manifold regime.
One also notes that in simplified models
without intermediate random manifold regime
(where one directly go form a Larkin regime to the asymptotic
regime) the above formula would give $a/R_a = (\pi c_L)^{2/(4-d)}$.
The transition then occurs for smaller values of $R_a/a$, in agreement with
the results found in a special geometry
\cite{kierfeld_bglass_layered,carpentier_bglass_layered}.

Using (\ref{lera}) and the expression (4.12) of 
\onlinecite{giamarchi_vortex_long} for $R_a$:

\begin{equation}
R_a = \frac{2 a^4 c_{66}^{3/2} c_{44}^{1/2}  }{
\pi^3 \rho_0^2 U_p^2 d  \xi^2}
\end{equation}

as well as $c_{66}=\epsilon_0/(4 a^2)$ 
and $c_{44} \approx c \epsilon_0 /(\gamma^2 a^2)$
(single vortex contribution) with $\epsilon_0 = (\Phi_0/4 \pi \lambda)^2$
and $c$ a numerical constant \cite{glazman_koshelev_decoupling}. One gets:

\begin{equation}
a^3 = \frac{4 \pi^3}{(\pi c_L)^6} 
\frac{U_p^2}{\epsilon_0^2}
d \xi^2 \frac{\gamma}{\sqrt{c}}
\end{equation}

One thus obtains an expression for the
transition field $H_M$ naturally expressed
in terms of some characteristic fields
of the system:
\begin{equation}
H_M(T=0) = \frac{(\pi c_L)^{4}}{(16 \pi)^{1/3} \pi^2}
(\frac{\epsilon_0}{U_p})^{4/3}
H_{c2}^{2/3} H_{\rm cross}^{1/3}
\end{equation}
where we have introduced the crossover field
$H_{\rm cross}=\pi c \Phi_0/(\gamma^2 d^2)$ with
$c \sim \ln(\gamma d/\xi)$ \cite{glazman_koshelev_decoupling}
and $H_{c2} = \Phi_0/\xi^2$.

As numerical estimate of the melting field $H_M$ for BiSCCO
with $H_{\rm cross} \sim 1T$, $H_{c2} \sim 100T$,
$U_p/\epsilon_0=0.4$, $c_L=.12$ gives
$H_M\sim 400 G$ in good agreement with the observed experimental values
\cite{khaykovich_diagphas_bisco}. The fact that this field
is well below the decoupling field validates a posteriori the
calculation ( note also that $B(r=0,z=d)$ is still small at the
transition). The general shape of the phase diagram
is in agreement with the one of Figure 1. Note that
some non linear effects, such as screening of 
disorder by thermal fluctuations or by interactions
at short scales may not be captured directly by the
gaussian theory of \onlinecite{giamarchi_vortex_long}.
They can be incorporated by a renormalization of 
the effective disorder $U_p(T)$. Such effects were
computed in the flux {\it liquid} using RG in \cite{nelson_ledoussal_liquid}
and it was shown that the pinning length was renormalized
upward (and thus the effective pinning strength downward)
by a factor $\exp((T/T_{dp})^3)$ where
$T_{dp} \sim (U_p^2 d \xi^2 c \epsilon_0/\gamma^2)^{1/3}$
is the single vortex depinning
temperature \cite{blatter_vortex_review}.
It would be interesting to compute these effects in the
solid as well. On general grounds that thermal fluctuations
can only weaken the disorder, one expects an 
additional curvature {\it upward} of the Bragg glass
instability line $H_M(T)$ when $T$ increases beyond $ O(T_{dp})$.

In Fig.1 two main regions can be distinguished:
if the temperature is high the stability line is nearly
indistiguishable from the melting line of the pure system. This regime
corresponds to the case where $R_a(T=0) \gg a$. In that case the
translational order is only affected at distances huge compared to the
lattice spacing, and the modification compared of a pure lattice is
negligeable as far as the melting is concerned. This part of the
stability line is therefore nearly identical to the melting of a pure
lattice and one can expect the transition to be first order. The Bragg
glass melts to a liquid phase,  nearly insensitive to disorder.

If the field is increased one will shorten $R_a(T=0)$. The disorder
itself is now able to make dislocation proliferate. In particular even
at $T=0$ disorder destroys the Bragg glass. In this range of field and
at low $T$ the
transition line flattens as a function of temperature, since it is controlled
mainly by the disorder. The phase into which the 
Bragg glass ``melts'' at low $T$ is
relatively poorly understood. It is characterized by an absence of
translational order and of Bragg peaks. There should still be
some amount of pinning at low $T$, but whether such a phase is a true glass
with diverging barriers, similar to the
proposed vortex glass of Ref. \cite{fisher_vortexglass_long},
or simply a very viscous form of the liquid phase
remains controversial. This phase could also retain hexatic 
order (hexatic glass) since, at least at a naive level,
similar arguments for survival of hexatic order as
for topological order in the Bragg glass can be given.
If the phase is a true glass phase then it should
melt thermally to the liquid, on the thick-dotted line of
Fig. 1. Whether a true vortex glass phase 
exists in {\it untwinned} samples is an important
still open and controversial question \cite{fendrich_irradiation_prl}
which may need to be settled by high
sensitivity \cite{charalambous_rings_prl} measurements.
Since the low temperature phase is in any case
much more continuously related to the liquid phase, one can expect now
the transition to become second order.
The Bragg glass therefore provides one natural explanation for a change of
the order of the (thermal) melting transition, as well as for the
existence of a field-induced transition.

One should also point out that at {\it very low} fields ($B \sim H_{c1}$)
where screening is important $a > \lambda$, a similar (inverted) field driven
transition should also occur when the field is lowered, from the
Bragg glass to a pinned liquid (or another glass) as suggested by the
decoration images of \cite{grier_decoration}. As shown in
\cite{nelson_ledoussal_liquid} the liquid becomes unstable
for $B < B_0 exp(-(T/T_{dp})^3)$, presumably to a glass or a pinned liquid.

The proposed field-induced transition between the Bragg
glass and the putative vortex glass being just characterized
by an injection of dislocations,
it is not necessarily linked to a decoupling
beween the layers. As a consequence one does not expect the critical
current along $z$ to become zero at the field-induced transition, at
least for low anisotropy system like YBCO. Of course it is always
possible that in materials with high anisotropy like BSSCO dislocations
prefer to appear first between the planes and the BG-VG and decoupling
transition coincide. Let us however emphasize that it does not need to
be so and that we also expect our transition to occur in purely
isotropic systems. Another argument against the field-induced transtion
being a simple thermal decoupling transition \cite{glazman_koshelev_decoupling},
is the fact that
such a transition could not extend down to zero temperature. In any case
measurements of the critical current perpendicular to the plane, in
particular in YBCO, should help to separate between the two effects.

The suggestion that there may be two different glass phases
could seem farfetched. There is a case however, mostly of theoretical
interest at present, where it should happen as a direct consequence
of our Bragg glass considerations. This is for $d$-dimensional vortex line
systems with correlated disorder or equivalently in $d-1$ dimensional
quantum bosons with disorder. It is reasonably well established
theoretically, numerically and experimentally that 
a Bose glass phase
exists for these systems in $d=3$ (i.e $d=2+1$ (2 space, 1 time dimension) for bosons).
It is also believed that this phase lacks translational long range
order in the plane
perpendicular to the columns. Indeed since the vortices are localized along
the columns, one can roughly view the properties in the
perpendicular plane perpendicular, as those of a $d-1$ dimensional
system with point-like disorder. For the $d=3$ vortex problem,
dislocations are therefore expected to appear (as they presumably appear for
$d=2$ systems with point like disorder). In higher dimension
however, this need not be the case. For instance in $d=4$ for vortex systems
($d=3+1$ for quantum particles) one is led, by similar arguments
as in \cite{giamarchi_vortex_long} to {\it two distinct localized phases}.
For weak disorder no dislocation will appear, giving a
Bose glass with topological order. This ``Bragg-Bose glass'' phase is
the equivalent for columnar defects of the Bragg glass one occuring for
point-like disorder. For stronger disorder, dislocations will destroy the
topological order perpendicular to the columns, giving back the
``conventional Bose glass'', i.e the continuation of its $d=2+1$ version.
At the transition between these two different Bose glass phases,
unbinding of dislocations loops (cylinders) should occur.
An interesting point is that in the ``conventional Bose glass''
these dislocations loops will remain pinned to the
colummnar defects, and it will thus be a true glass. This
Bose glass phase would be in some sense analogous, in the 
case of point disorder, to the putative vortex glass.
However, for pointlike disorder it is much less obvious that
such a phase exists as a genuine thermodynamic phase. Let us note that
this problem was studied analytically within an elastic theory
in \cite{giamarchi_variational_bose} using a
variational method and in \cite{balents_rg_bose} using
RG methods: the phase described there is thus 
the ``Bragg-Bose glass''. The difference between the two phases
should occur only at scales larger than the distance
between unpaired dislocations.

Several consequences of our theory could be further checked in
experiments. First since the Bragg glass phase has translational order and the
vortex glass
has not, neutron experiments should observe a destruction of the Bragg
peaks at the same location
\cite{footnote3}
than the transition observed by magnetic
measurements.  Such a feature seems to be consistent with the
existing experimental data in BSSCO \cite{forgan}, but a
more detailed experimental investigation would be needed to check
this point in other materials as well.
Another clear distinction between the two phases should be observed when
a cycling in current similar to the one of \cite{yaron_neutrons} is
performed. Such cycles taking the system above the critical $J_c$ and
then back to zero, are expected to heal the lattice and to expel out of
equilibrium dislocations. One can therfore expect good healing in the
Bragg glass phase, as is indeed the case, since dislocations can only
exist as out of equilibrium object. On the other hand the same cycle
performed in the VG phase should make little difference on the neutron
diffraction pattern since the {\it equilibrium} state
already contains unpaired dislocations.

\begin{figure}
\label{figure2}
\centerline{\epsfig{file=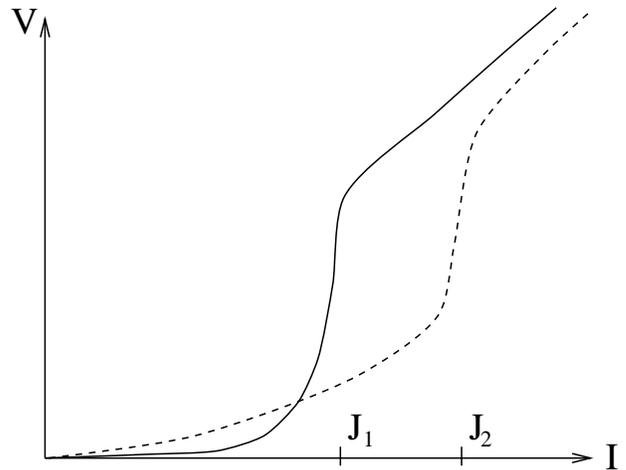,angle=-90,width=8cm}}
\vspace{0.5cm}
\caption{$I-V$ characteristics in the Bragg glass phase and in the
vortex
glass (or pinned liquid) are shown
schematically respectively as solid and dashed lines. One goes from
small ($J_1$) to larger ($J_2$) critical currents,
but rapidly divergent to weakly
divergent (or finite) barriers, when increasing the field.}
\end{figure}

Finally one expects the barriers to vary very differently in the two
glass phases. In the Bragg glass phase, elasticity is strong. Pinning
can only be collective and one expects therefore weak barriers at short
length scales. This implies a small critical current. On the other hand,
since creep can only occur collectively, the barrier should grow very
rapidly with decreasing current. A simple estimate
gives $U(j) \sim (1/j)^\mu$ with $\mu$ ranging between
$\mu=0.7-0.8$ at intermediate currents to $\mu=0.5$ at small currents
\cite{giamarchi_vortex_long}
while taking the dispersion of elastic moduli into account leads
to higher values of $\mu$ in the intermediate regime
\cite{blatter_vortex_review}. On the other hand in the VG phase barriers
should be significantly larger at
short length scales since the nearly destroyed lattice has additional
effective degrees of freedom, such as free dislocations, and can thus
adapt more easily to the pinning potential.
The critical current should therefore
increase when approaching the field-induced transition.
On the other hand in the VG phase the barriers should grow much
more slowly with decreasing current since there is no need for
collective motion, or even remain finite if the phase
is simply a crossover from the liquid phase. Some estimates of the
exponents for the gauge glass model gave very small exponents of the
order of $\mu \sim 0.1-0.2$. One can therefore expect $I-V$
characteristics evolving
with fields like the ones shown in Figure~2.
Such a behavior is in good qualitative agreement of the observations of
a second peak in relaxation measurements when the transition is passed.
More refined transport, or relaxation measurement should help deciding
on the behavior of the barriers.

It is important to note that the Lindemann criterion used here
is {\it not} a detailed theory of the transition when dislocation proliferate,
which is not yet available. It represents one possible mechanism 
of instability dominated by short length scales. 
It thus provides a reasonable upper bound for the instability field $H_M$ since
the Bragg glass certainly cannot self-consistently survive if $R_a < a$.
However, it cannot be excluded that because of the weakening 
of translational order at large distances in the Bragg glass
compared to a real solid, unbound dislocations start to appear first
at large length scales providing a different instability mechanism.
In that case this additional phase (which may or may
not be a true glass) would also melt through a first order
transition with good short distance translational order properties.

In conclusion, we have examined in some details the implications, 
for the phase diagram of type II superconductors,
of the existence of a glass phase with translational order: the Bragg glass.
We have shown that it provides a natural interpretation of many of the
features of the phase diagram of BSCCO and YBCO observed in experiments.
Namely a crossover from a first order melting
transition to a continuous transition when the field is increased, and the
existence of a field induced transition. We interpret this last
transition as the destruction of the Bragg glass phase by spontaneous
injection of unbounded
dislocations, into a topologically disordered glass phase or liquid. 
This transition, being disorder driven, should extend down to $T=0$.


\begin{thebibliography}{10}

\bibitem[*]{junk}
Laboratoire Associ\'e au CNRS. email: giam@lps.u-psud.fr.

\bibitem[**]{frad}
Laboratoire Propre du CNRS, associ\'e \'a l'Ecole Normale Sup\'erieure
et \`a l'Universit\'e Paris-Sud. email: ledou@physique.ens.fr.

\bibitem{blatter_vortex_review}  G. Blatter {\it et~al.},
Rev. Mod. Phys. {\bf 66}, 1125 (1994).

\bibitem{nelson_melting}  
D.R. Nelson, Phys. Rev. Lett. {\bf 69} 1973 (1988).

\bibitem{subdo_melting}
A. Houghton, R.A. Pelcovits and A. Sudbo,
Phys. Rev. B {\bf 40} 6763 (1989).

\bibitem{melting_experiments}
P.L. Gammel, L. F. Schneemeyer, J.V. Waszcak and D. J. Bishop
Phys. Rev. Lett. {\bf 61} 1666 (1988).

\bibitem{fisher_vortexglass_short}
M.~P.~A. Fisher, Phys. Rev. Lett. {\bf 62},  1415  (1989).

\bibitem{feigelman_collective}
M. Feigelman, V. Geshkenbein, A. Larkin,
and V. Vinokur, Phys. Rev. Lett.
{\bf 63},  2303  (1989).

\bibitem{nelson_ledoussal_liquid}
D.R. Nelson and P. Le Doussal, Phys. Rev. B {\bf 42} 10113 (1990).

\bibitem{fisher_vortexglass_long}  D. S. Fisher and M. P. A. Fisher and D.
A. Huse, Phys. Rev. B {\bf 43} 130 (1991).

\bibitem{chudnovsky} E. Chudnovsky, Phys. Rev. Lett. {\bf 65} 3060 (1990).

\bibitem{bouchaud_variational_global}
J. P. Bouchaud, M. M{\'e}zard, and J. Yedidia, Phys. Rev. Lett. {\bf 67},
3840 (1991) and Phys. Rev. B {\bf 46},  14686
(1992)

\bibitem{charalambous_melting_rc}
M. Charalambous, J. Chaussy, and P. Lejay, Phys. Rev. B {\bf 45},  5091
  (1992).

\bibitem{safar_tricritical_prl}
H. Safar {\it et~al.}, Phys. Rev. Lett. {\bf 70},  3800  (1993).

\bibitem{grier_decoration}
D. Grier et al. Phys. Rev. Lett. {\bf 66} 2270 (1991).

\bibitem{villain_cosine_realrg}
J. Villain and J.~F. Fernandez, Z Phys. B {\bf 54},  139  (1984).

\bibitem{nattermann_pinning}  T. Nattermann, Phys. Rev. Lett. {\bf 64}, 2454
(1990).

\bibitem{giamarchi_vortex_short}
T. Giamarchi and P. {Le Doussal}, Phys. Rev. Lett. {\bf 72},  1530  (1994).

\bibitem{giamarchi_vortex_long}
T. Giamarchi and P. {Le Doussal}, Phys. Rev. B {\bf 52},  1242  (1995).

\bibitem{footnote1} 
In \cite{giamarchi_vortex_long} we have computed
the quantity $B(r)$ using the RG and a variational method.
To then estimate the decay of translational order correlation
function $C_K(r)=\langle \exp(i K.(u(r)-u(0))) \rangle$
we have used that $C_K(r) \approx  \exp( - K^2 B(r)) \sim 1/r^{A_d}$,
i.e a gaussian approximation. This is
a reasonable lower bound for $C_K(r)$. It may give
the exact asymptotic decay or it is also possible that atypical ``return to
the origin'' events (i.e a singularity at $u=0$ of the scaled probability of $u$)
could make this decay {\it slower}.
A similar situation is discussed in \cite{mitra_pld_rmn}.


\bibitem{mitra_pld_rmn}
P.P. Mitra, P. Le Doussal, Phys. Rev. B  {\bf 44} 12035 (1991).

\bibitem{footnote2}
One can in principle use a more realistic correlator
(such as one with algebraic decay), but this should not change
the general form of our results.

\bibitem{gingras_dislocations_numerics} M. J. P. Gingras and D. A. Huse
Phys. Rev. B {\bf 53} (1996).

\bibitem{ryu_diagphas_numerics} S. Ryu, A. Kapitulnik and S. Doniach
Phys. Rev Lett. 77 (1996). For
an early discussion of the possible phase diagram see also
A. Kapitulnik et al. SPIE proceedings 2157 12 (1994).

\bibitem{kierfeld_bglass_layered} J. Kierfeld, T. Nattermann,  T. Hwa
preprint cond-mat/9512101, to be published.

\bibitem{carpentier_bglass_layered} 
D. Carpentier, P. Le Doussal
and T. Giamarchi, Europhys. Lett. {\bf 35} 379 (1996)
and to be published.

\bibitem{kwok_irradiations}
W. Kwok et al. Physica B {\bf 197}
579 (1994)

\bibitem{yaron_neutrons}
U. Yaron et al. Phys. Rev. Lett. {\bf 73} 2748 (1994)

\bibitem{giamarchi_comment_neutrons}  T. Giamarchi and P. Le Doussal Phys.
Rev. Lett. {\bf 75} 3372 (1995).

\bibitem{forgan}
R. Cubbit et al. Nature {\bf 365} 407 (1993),
Forgan et al. Preprint (1996).

\bibitem{melting_bscco}
E. Zeldov et al. Nature {\bf 375} 373 (1995).

\bibitem{khaykovich_zeldov} B. Khaykovich et al.
Phys. Rev. Lett. {\bf 76} 2555 (1996).

\bibitem{khaykovich_diagphas_bisco}
B. Khaykovich et al. preprint (1996).

\bibitem{glazman_koshelev_decoupling}
L.~I. Glazman and A.~E. Koshelev, Phys. Rev. B {\bf 43},  2835  (1991).

\bibitem{larkin_ovchinnikov_pinning}
A.~I. Larkin and Y.~N. Ovchinnikov, J. Low Temp. Phys {\bf 34},  409  (1979).

\bibitem{footnote3} How fast should one see the Bragg peaks 
disappear in experiments is
unclear. The presence of disorder and temperature induced
{\it finite} dislocation loops as a
precursor near $H_M$ could make the neutron peaks disappear
more gradually. We thank T. Nattermann for a discussion on
this point.

\bibitem{giamarchi_variational_bose}  T. Giamarchi and P. Le Doussal
Phys. Rev. B {\bf 53} 15206 (1996).

\bibitem{balents_rg_bose} L. Balents, Europhys. Lett. {\bf 24} 489 (1993).

\bibitem{fendrich_irradiation_prl}
J. A. Fendrich et al. Phys. Rev. Lett. {\bf 74} 1210 (1995).

\bibitem{charalambous_rings_prl}
M. Charalambous et al., Phys. Rev. Lett. {\bf 75} 2578 (1995).

\bibitem{safar_transport_tricritical}
H. Safar et al. Phys. Rev. B {\bf 52} 6211 (1995).


\end{thebibliography}

\end{multicols}
\end{document}